# Design of Recognition and Evaluation System for Table Tennis Players' Motor Skills Based on Artificial Intelligence

Zhuo-yong Shi[a], Ye-tao Jia[a], Ke-xin Zhang[b], Ding-han Wang[a], Long-meng Ji[a], and Yong Wu[a]

[a] School of Electronics and Information, Northwestern Polytechnical University, Xi'an 710129, China

[b] School of Foreign Studies, Northwestern Polytechnical University, Xi'an 710129, China

**Abstract**：With the rapid development of electronic science and technology, the research on wearable devices is constantly updated, but for now, it is not comprehensive for wearable devices to recognize and analyze the movement of specific sports. Based on this, this paper improves wearable devices of table tennis sport, and realizes the pattern recognition and evaluation of table tennis players' motor skills through artificial intelligence. Firstly, a device is designed to collect the movement information of table tennis players and the actual movement data is processed. Secondly, a sliding window is made to divide the collected motion data into a characteristic database of six table tennis benchmark movements. Thirdly, motion features were constructed based on feature engineering, and motor skills were identified for different models after dimensionality reduction. Finally, the hierarchical evaluation system of motor skills is established with the loss functions of different evaluation indexes. The results show that in the recognition of table tennis players' motor skills, the feature-based BP neural network proposed in this paper has higher recognition accuracy and stronger generalization ability than the traditional convolutional neural network.

# Introduction

Sports have the functions of physical fitness and recreation, and they are indispensable in the growing process of teenagers. More and more attention has been paid to the positive role of sports in promoting health, which is followed by the continuous research on sports [1-5]. With the rapid development of electronic information, intelligent wearable devices are increasingly widely applied in sports [6-8]. Intelligent wearable devices can be used to monitor athletes' physical signs and parameters, and help us better do physical exercise [9-11].

Relevant scholars' researches on intelligent wearable devices mainly focus on the directions of low power consumption, wear resistance, radiation, privacy, reliability and device lightweight, etc. In terms of function, they mainly realize the calculation of motion characteristics or the estimation of exercise calories consumption. Relevant scholars have also achieved a lot of achievements in their research on the performance of various aspects of wearable devices. For example, Wu[12] et al. proposed that wearable devices should design low-power system or apply energy harvesting technologies, in order to minimize interpersonal interaction and enable wearables to run for a long time without changing batteries or recharging. Chen[13] et al. introduced the idea of artificial integration, which attempted to solve the discomfort caused by wearing many sensors in different parts of the body for healthcare applications. All wearable Internet of Things (IoT) devices use wireless technology to transmit their sensing data to another node, gateway or base station. This wireless transmission involves radio-frequency radiation, which can have a negative impact on the user's health. Dian[14] et al. solved this safety problem by analyzing the standard limit of human exposure to radio-frequency electromagnetic energy and the radiation level of CIoT antenna. Zhao[15] et al. proposed that smart glasses often present information on a screen very close to people's eyes, and long-term use may bring side effects such as headache, dizziness and other discomfort. In addition, since the sensitive data collected by smart wearable devices is easy to be stolen, the privacy protection of user data of wearable devices is also very important. Lee[16] et al. showed the privacy risks caused by keystroke inference attacks. RK[17] et al. proposed a broadcast user IoT model and put forward a healthcare facility or user-authorized device in which users' personal data is only shared with expected nodes. What's more, some special equipment reliability is also very important. UMAMAHESWARAN[18] et al. designed a body temperature sensor, pulse monitoring sensor, Global Positioning System (GPS) tracker, the display with an output values, communication service network supported by the IoT and battery for operating the sensor. He set up preventative tools to notify relatives and doctors of sensor outliers. Due to the portable nature of wearables, lightweight data is also important. Yoon[19] et al. proposed a new concept of lightweight user interaction. They used ready-made ambient light sensors as technology substitutes for user interfaces of smart wearable devices, and designed and modeled lightweight user interaction according to typical uses of representative smart wearable devices.

In the field of table tennis, the motor skills of table tennis players often rely on the coach's subjective evaluation of the corresponding motor skills, and this artificial evaluation depends very much on the professional level of the coach. Therefore, an objective evaluation system is needed to eliminate the subjective influence of manual evaluation, and the emergence of wearable devices makes it possible. However, current wearable devices are more friendly to general sports, but the motion pattern recognition and analysis of specific sports are not comprehensive [20]. It only measures and calculates the steps and calories of the players, which cannot meet the needs of table tennis players. In table tennis, the players are more concerned about the use of all the motor skills and the standardization of motor skills in a table tennis game, which requires the development of a wearable device that can be used to identify and analyze the motor skills of table tennis players.

The use and the standardization of various motor skills of table tennis players come down to the study of pattern recognition, which was first widely used in the field of aeronautics and astronautics to monitor the flight status of electronic equipment in space. With the rapid development of pattern recognition, there are more and more application in other fields. For example, in the field of biology, the modeling of complex macromolecules [21], the identification of microbial characteristics [22], and the screening of dissociated cells [23]; In computer applications, retrieval feature syntax [24], face recognition system [25], character recognition system [26]; Recognition of acquired dermatosis in medical field [27], application of brain-computer interface [28], cancer treatment [29]; In addition, pattern recognition has also been applied in other fields, such as instrument fault recognition [30], image processing [31], agricultural product quality supervision [32], etc. Pattern recognition is a process of describing, identifying, classifying and explaining things or phenomena by processing and analyzing various forms of information that represent things or phenomena. Current researches on pattern recognition by relevant scholars include manipulator control, facial expression recognition, EEG emotion recognition, pathogen recognition, pattern recognition of disease-resistant enzymes, etc. Lu[33] et al. designed a manipulator control, which enables stroke patients to use pattern recognition to control manipulator control activities for rehabilitation training of patients. Rivera[34] et al. divided the face into several regions and adopted the local directional pattern to improve the accuracy of facial features and expression recognition. Li[35] et al. used power spectrum analysis method to improve the performance of emotion recognition by integrating EEG features with brain information transmission patterns. Saijo[36] et al. demonstrated the dynamic and convergent evolution of pattern recognition specificity and proposed the functional implications of pathogen recognition and disease resistance derived from pattern recognition receptors. Seo[37] et al. designed a pattern recognition device based on optical synapses, which can simulate the color and color mixing pattern recognition function of human visual system when it is arranged in optical neural network. Boutrot[38] et al. proposed to detect plant insect pests through pattern recognition receptors that produce disease-resistant enzymes in plants, and that this technology has potential applications in emerging biotechnologies that improve crop broad-spectrum and potentially long-lasting disease resistance.

With the rise of artificial intelligence and the substantial improvement of computer computing power, it is possible to use artificial intelligence technology to realize the use of various motor skills and the standard degree of motor skills in the process of table tennis players. Based on the research demand of wearable devices for recognition and analysis of table tennis players' motor skills, this paper study the use of various motor skills in table tennis and the standardization of motor skills though artificial intelligence. This research is mainly divided into two parts: 1) The pattern recognition of various motor skills of table tennis players. 2) Evaluation system of table tennis players' motor skills. The key points are as follows:

1) The following elements are included in the pattern recognition of motor skills. Firstly, the window function is used to segment the collected signals. Secondly, multi-dimensional features of table tennis players are constructed based on feature engineering. Thirdly, the dimensionality of the features is reduced based on PCA method. Finally, the athletes' motor skills are identified and predicted based on neural network model and support vector machine model respectively.

2) In the evaluation system of motor skills, the first is establishing the motor skills evaluation system based on multiple evaluation indexes. The second is calculating the evaluation index data of collected data and then the distance between evaluation indexes at all levels and standard indexes. At last, the motor skills of athletes are scored.

The rest of this article is organized as follows. The first part is about the collection and preprocessing of table tennis players' sports data. It introduces the data acquisition system built in this study and the contents of outlier removal, missing item interpolation and high-frequency noise filtering. In the second part, the sliding window function is used to realize the data segmentation of table tennis players' movement data and the database establishment of the features of table tennis standard motor skills. The third part is the recognition model of table tennis motor skills. Based on feature engineering, multi-dimensional features are constructed and effective features are extracted, which are used to classify and recognize table tennis movements. The fourth part is the establishment of table tennis motor skill evaluation system including motor skill hierarchical evaluation system and the loss function of different evaluation indexes. The fifth part is about the prediction results of the two classification models and compares them with the traditional models. Finally, the sixth part is the summery of this paper.

# 1. Data collection and preprocessing

## 1.1 Design of data collection device

The information collection device of table tennis movement is composed of power module, gyroscope module, accelerometer module and CPU. The movement information collection system built in this study is shown in Figure 1. The Power system, Gyro module and Accelerometer module of the lower computer of table tennis movement information collection system are controlled by the CPU, and communicate with the upper computer PC through the transmission system so as to realize the data collection of the system.

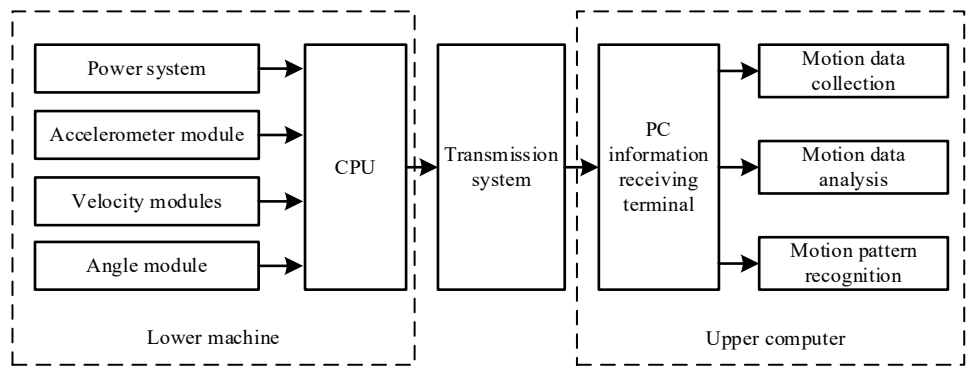

Figure 1. Structure Diagram of Table Tennis Movement Information Collection System

As Figure 2 shows, the hardware diagram of table tennis movement information collection system is built according to Figure 1. In Figure 2, there are power supply modules, buck modules, master control module, Bluetooth modules, posture module and display module are respectively included in the table tennis movement information collection system. The power supply modules consist of two batteries in series, which can provide 7.4V-8.4V voltage and power resources. The buck modules are composed of 2596HV buck module, which can stabilize the input voltage to 5V, and stably supply power to the master control system. The master control module is STM32F103ZET6, which can realize the control and transmission of data and can be used to collect, display and send data. The Bluetooth modules include HC-06 module, which can realize Bluetooth wireless communication and is used for

Bluetooth wireless transmission of data. The posture module is of MPU6050, which can realize the perception of spatial posture and collect the corresponding posture information. The display module is OLED module, which can realize the display of 128*64 resolution to show the collected posture information.

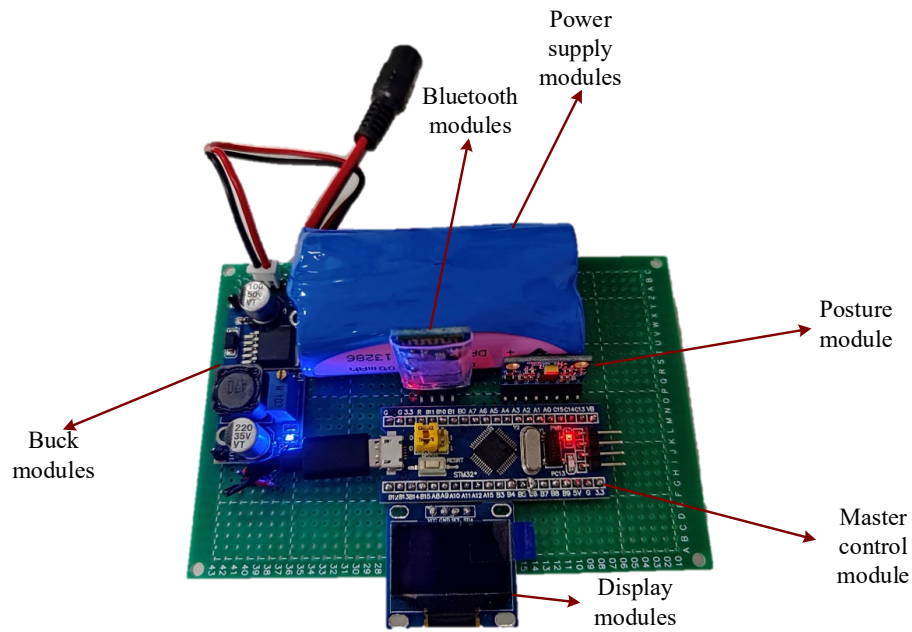

Figure 2 Hardware Diagram of Table Tennis Movement Information Collection System

## 1.2 Data collection

Players need to decide the direction, velocity, strength and placement of the return ball according to their position and their own skills, and choose six kinds of motor skills in the process of return, that is, forehand/backhand attack, forehand/backhand push and forehand/backhand chop. In this experiment, triaxial acceleration data, triaxial angular velocity data and triaxial Angle data are collected during motion. And the data is transmitted in the form of vector set, which is: $<acc, \omega, \theta>$, $acc$ including $<acc\_x_i, acc\_y_i, acc\_z_i>$, $\omega$ including $<\omega\_x_i, \omega\_y_i, \omega\_z_i>$, $\theta$ including $<\theta\_x_i, \theta\_y_i, \theta\_z_i>$. The athletes' movement data is collected through the table tennis movement information collection system as shown in Figure 2, and the movement data collected is shown in Table 1.

| Table 1 List of collected data | | |
|---|---|---|
| Sample Name | Sample | Interval |
| Forehand Attack Data | 100000 | 0.01s |
| Backhand Attack Data | 100000 | 0.01 s |
| Forehand Push Data | 100000 | 0.01 s |
| Backhand Push Data | 100000 | 0.01 s |
| Forehand Chop Data | 100000 | 0.01 s |
| Backhand Chop Data | 100000 | 0.01 s |

Taking the information collected within 10 seconds as an example, Figure 3 shows the collected data of table tennis players in their sport process.

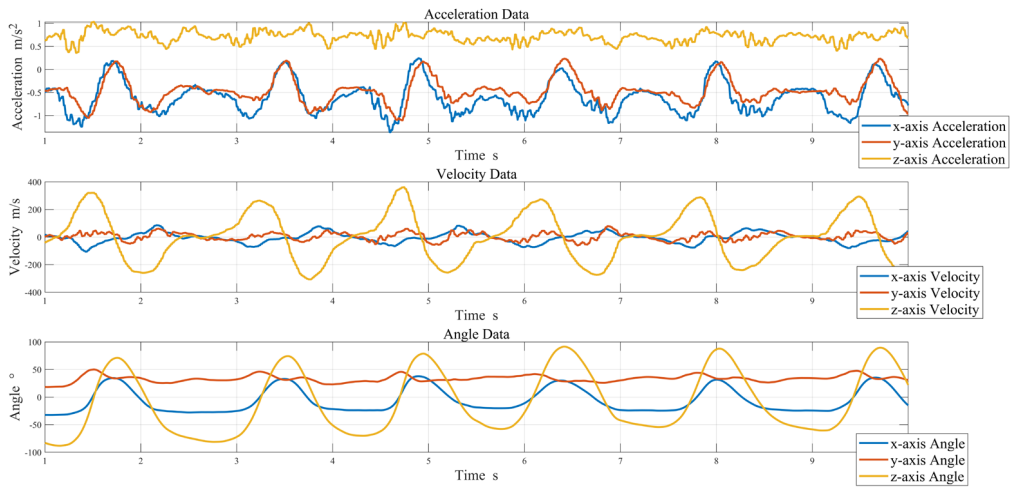

Figure 3 The Collected Data of Table Tennis Players in the Sport Process

## 1.3 Data cleaning

After collecting the acceleration data of table tennis players in the process of movement through the device in 1.1, there are some abnormal data in the data, so it is necessary to carry out data cleaning. The data cleaning in this study mainly includes the removal of abnormal data and interpolation of missing data.

1.3.1 The removal of abnormal data

According to a group of n samples of sports data of table tennis players $\{x_1, x_2, \ldots, x_n\}$, the first-order difference $X_i$ is calculated as shown in the equation.

$$X_i = x_{i+1} - x_i \ (i = 1, 2, \ldots, n-1) \tag{1}$$

$X_i$ is the first-order difference of motion data collection, which describes the changes of motion

data collection in a period of time. According to statistical knowledge, the changes should follow normal distribution, so statistical characteristics of $X_i$ are collected for data removal.

Firstly, the mathematical expectation $EX$ of the first-order difference data $X_i$ of the motion sample $x_i$ is calculated as shown in the equation.

$$EX = \frac{1}{n-1} \sum_{i=1}^{n-1} X_i \tag{2}$$

Then, the standard deviation $\sigma$ of the first-order difference data $X_i$ of the motion sample $x_i$ is shown in the equation.

$$\sigma = \sqrt{\frac{1}{n-1} \sum_{i=1}^{n-1} (X_i - EX)^2} \tag{3}$$

On the grounds of the principle of statistics, the first-order difference value $X_i$ of the collected motion data sample $x_i$ should follow the normal distribution. According to the $3\sigma$ criterion, the data within $(EX - 3\sigma, EX + 3\sigma)$ should be retained as normal data, and the data outside $(EX - 3\sigma, EX + 3\sigma)$ should be excluded as abnormal data. The abnormal data $X_i$ and the corresponding data $x_{i+1}$ should be removed, so as to realize the removal of abnormal data.

1.3.2 The interpolation of missing data

After the removal of abnormal data in 1.3.1, the data in the deletion position needs to be repaired by interpolation. In this study, the Newton interpolation method is used for recovery of the missing data, and then to output the recovered data.

For a group of data $(x_i, f(x_i))(i = 1 \cdots n)$ after the removal of outliers, $x_i$ represents the position of the independent variable, and $f(x_i)$ is the function value of this position. Every $x_i$ is different, and the missing data is denoted as $x_j$, and the closest four data of $x_j$ are $(x_0, f(x_0))$, $(x_1, f(x_1))$, $(x_2, f(x_2))$, $(x_3, f(x_3))$, which are used for data recovery.

The first step is to define the first-order difference quotient between any two nodes as shown in the equation.

$$f[x_i, x_j] = \frac{f(x_j) - f(x_i)}{x_j - x_i} \tag{4}$$

In the equation, $f(x_j)$ is the function value of $x_j$, and $f[x_i, x_j]$ is the first-order difference quotient of $x_i$ and $x_j$.

Secondly, the k-order difference quotient of k+1 non-identical nodes is defined as shown in the equation.

$$f[x_0, x_1, \cdots x_k] = \frac{f[x_1, x_2, \cdots x_k] - f[x_0, x_1, \cdots x_{k-1}]}{x_k - x_0} \tag{5}$$

In the above equation, $f(x_j)$ is the function value of $f(x_j)$, $f[x_0, x_1, \cdots x_{k-1}]$ is the first-order difference quotient of $x_0, x_1, \cdots x_{k-1}$.

A difference quotient table is established for known data, and based on the difference quotient table, the Newton interpolating polynomial $N_3(x)$ is constructed at $x_j$ is shown in the following equation.

$$N_3(x) = f(x_0) + f[x_0, x_1]\omega_1(x) + f[x_0, x_1, x_2]\omega_2(x) + f[x_0, x_1, x_2, x_3]\omega_3(x) \tag{6}$$

There is $\omega_i(x)$ as the group of interpolation basis functions, which can be represented as follows.

$$\omega_i(x) = \prod_{k=1}^{i}(x - x_k) \tag{7}$$

The interpolation remainder of the interpolation polynomial is shown in the next equation

$$R_3(n) = \frac{f^{(n+1)}(\zeta)}{(n+1)!} \tag{8}$$

The above interpolation method can realize the recovery of missing data, and the function value of recovered data is shown in the equation.

$$f(x_j) \approx N_3(x_j) \tag{9}$$

A third-order difference quotient table was constructed based on the Newton interpolation method to recover the function values of missing data, and to realize the data interpolation estimation where the data is missing .

## 1.4 High-frequency noise filtration

There is certain high-frequency noise in the collection process of table tennis movement data. Based on this, the adaptive smoothing filtering method is adopted to process the data. In order to adapt to the real-time performance of the signal, the filtering parameter is updated by increasing the threshold.

The new output value is defined as a linear combination of the current sampled value and the last output value. The filter output is shown in the equation.

$$Y(n) = m \cdot X(n) + (1-m) \cdot Y(n-1) \quad (10)$$

In the above equation, $Y(n)$ is the filtering output value of the filter. $m$ is the filtering coefficient between [0,1]. $X(n)$ is the current sampled value. $Y(n-1)$ is the output value of the previous filtering.

As can be seen from the above-mentioned equation, the filtering coefficient is the most important parameter of the filter, which has a great influence on the filtering effect. In order to make the filter design more reasonable, the time-variant features of the input signal are tracked and the filtering parameters are updated during the output of the signal. The conditional threshold $\Delta a$ for determining motion state is introduced, and the filtering coefficient $m$ is judged and modified according to the equations shown in the following equation.

$$\begin{cases} \Delta = Y(n) - Y(n-1) \\ m = \left(1 - \dfrac{\Delta_a}{\Delta}\right) \cdot k_0 \end{cases} \quad (11)$$

As shown in this equation, $\Delta$ is the difference between this filtering output value and the output value of the last filtering. $\Delta_a$ is the conditional threshold for determining the motion state, and $k_0$ is the default filtering parameter.

The movement data collection device of table tennis players in 1.1 is used to collect the movement data of table tennis players. The acceleration data in the x direction in the movement data of a group of table tennis players is taken as an example for data preprocessing according to the methods in 1.3 and 1.4. The processing effect is shown in Figure 4.

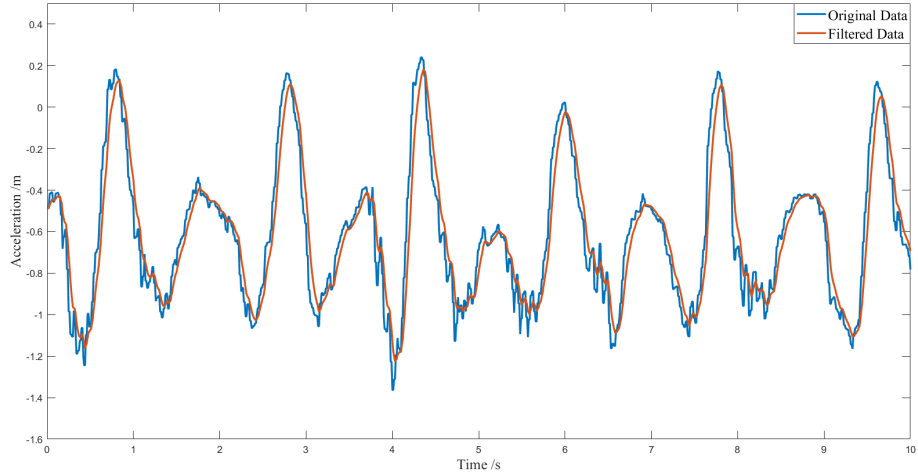

Figure 4 The Image of Data Preprocessing

In Figure 4, the acceleration data in the X-axis direction during athletes' movement is taken as an example to verify the effect of data preprocessing. The sampling period is 0.01s, and 10 seconds of acceleration data are collected for data preprocessing. It can be seen from the figure that the error signals generated in the collection process can be well filtered after data preprocessing, which realizes the expectation of data smoothing.

## 2. Establishment of standard motor skill models

The benchmark motor skills of table tennis mainly include forehand attack, forehand push, forehand chop, backhand attack, backhand push, backhand chop, etc. The six standard motor skill models are established by collecting motor skill data of standard table tennis players. Based on these models, the loss evaluation index of each motor skill is given to measure the loss of any part of motor skills and standard motor skills.

### 2.1 The design of sliding window function

The athletes' movement information collected is transmitted in the way of data stream and cannot be processed directly. Therefore, it is necessary to segment the collected data. The common solution in data partition is to set a sliding window for processing the collected data. The data collected and processed by windows can be divided into corresponding data segments according to the window function, which provides the basis for the processing of the following feature engineering. Relevant scholars have conducted many experimental studies on sliding window function and found that setting the overlap rate of window function as 50% is a feasible method [39,40]. The window setting of the window function will directly affect the recognition of movements. If the window setting is too large, it may cause delay of movement recognition or training errors by recognizing various types of movements. If the window function is too small, it may not recognize a whole movement, thus introducing identification errors .

After a comprehensive statistical analysis of the time required by each movement in table tennis, 2s is selected as the default duration of a movement, that is, a sliding window. In combination with the sampling frequency of the movement data in this study, which is 100Hz, the width of the sliding window is set as 200 sampling points, and the coincidence rate of the sliding window is 50%, as shown in Figure 5. By setting the sliding window function, the length of different movement data can be consistent, which is convenient for the statistics of movement data and provides significant help for the following feature extraction and movement recognition.

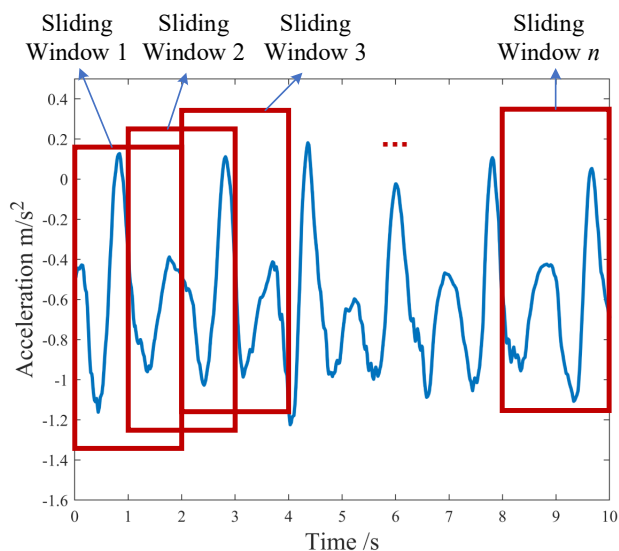

Figure 5 The Movement Data Segmentation of Sliding Window Function

## 2.2 The design of motor skill activation function

In the collected data of table tennis players, there can be a situation without table tennis movements, which causes that the data divided through the sliding window does not represent the table tennis motor skill, and this kind of data will trouble the motor skill recognition. Therefore, after segmenting the sliding window function data, it is necessary to determine whether the players do motor skills in this group of sliding windows. In order to solve this problem, the activation function of motor skills is designed to judge whether a table tennis player completes a whole motor skill in a sliding window.

The information of acceleration and angular velocity in table tennis movements is used as the marker to judge whether there is an effective motor skill in the corresponding sliding window. Common binary classification models include logistic regression, naive Bayes, decision tree, support vector machine, random forest, etc. Considering that table tennis motor skill activation function is not a set of nonlinear and complex classification, the support vector machine model is thus selected for this classification. The support vector machine model is supported by rigorous mathematical theory, and a unique hyperplane can be determined by support vector to classify data. The model is of strong interpretation and high stability.

In the support vector machine, for a given binary data set $D = \{(x_i, y_i)\}_{i=1}^{N}$ and $y_i \epsilon \{+1, -1\}$, if two samples are linearly separable, that is, there is a hyperplane $\omega^T x + b = 0$ separating two types of samples, then for each sample, $y_i(\omega^T x + b) > 0$. At the same time, considering that the model needs to have certain error information and prevent model from overfitting, the optimization problem of both empirical risk and structural risk is considered in the objective function. Therefore, the optimization equation of this problem is simplified as shown in the equation.

$$\min \|\omega\|^2 + C \sum_{i=1}^{n} \xi_i \qquad (12)$$

In the above equation, $\omega$ is interval. $\xi_i$ is slack variable. $C$ is weight coefficient, which is used to determine the weight between empirical risk and structural risk.

The constraints of the optimization equation are shown in the equation.

$$s.t. \begin{cases} y_i(\omega^T x + b) \geq 1 - \xi_i \\ \xi_i > 0 \end{cases} \qquad (13)$$

In the above equation, $y_i$ is sample label, $\omega$ is interval, and $\xi_i$ is relaxation variable.

The activation function of motor skills is determined by the optimization model as shown in the equation in order to decide whether there are effective movements in the current sliding window. When the motor skill activation function determines that the motor skill data in the sliding window is effective, the subsequent method is adopted to identify the motor skill data. When the action skill activation

Function determines that the movement data in the sliding window is not effective, this set of data is skipped and the next data in the sliding window is carried out to design the movement activation function.

## 2.3 The establishment of standard movements

In table tennis, common benchmark motor skills mainly include forehand attack, forehand push, forehand shop, backhand attack, backhand push, backhand chop, etc. The basic six motor skills are shown in Figure 6. M1-M6 describes the six movements of forehand attack, backhand attack, forehand push, backhand push, forehand chop and backhand chop respectively. The sampling period is 10ms, and the standard period is 2000ms.

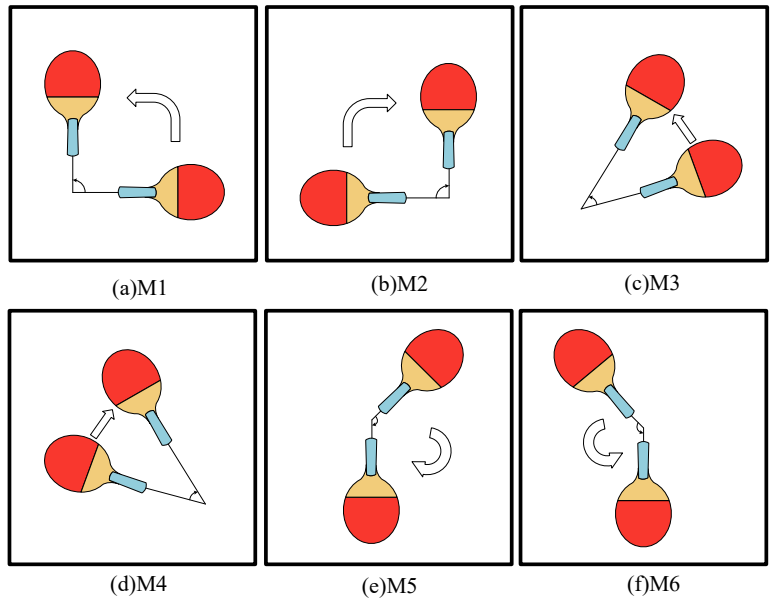

Figure 6  Six Benchmark Motor Skills in Table Tennis

Various benchmark motor skills in Figure 4 are collected for 50 periods, and the feature database of six motor skills is established, as shown in Figure 7.

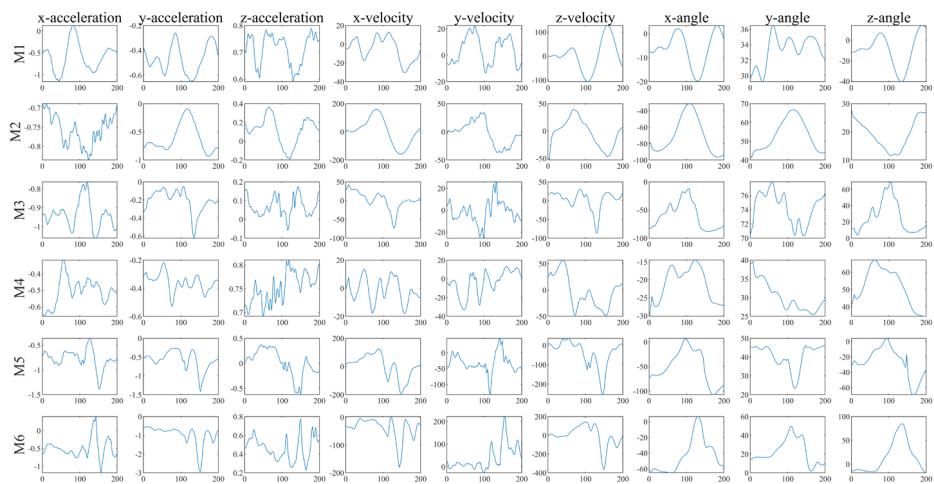

Figure 7 The Feature Database of Six Motor Skills

# 3 The model of motor skill recognition

Table tennis mainly includes forehand attack, forehand push, forehand chop, backhand attack, backhand push, backhand chop. In this chapter, the model of table tennis motor skill recognition is established based on two classification methods respectively, by which six benchmark movements of table tennis are classified, so as to realize the recognition of table tennis players' motor skills. The table tennis player movement recognition and evaluation algorithm proposed in this paper is shown in Figure 8. In this algorithm, the multilevel index constructed based on feature engineering realizes the classification and recognition of table tennis motor skills by using BP neural network and the hierarchical evaluation index evaluation system based on feature database.

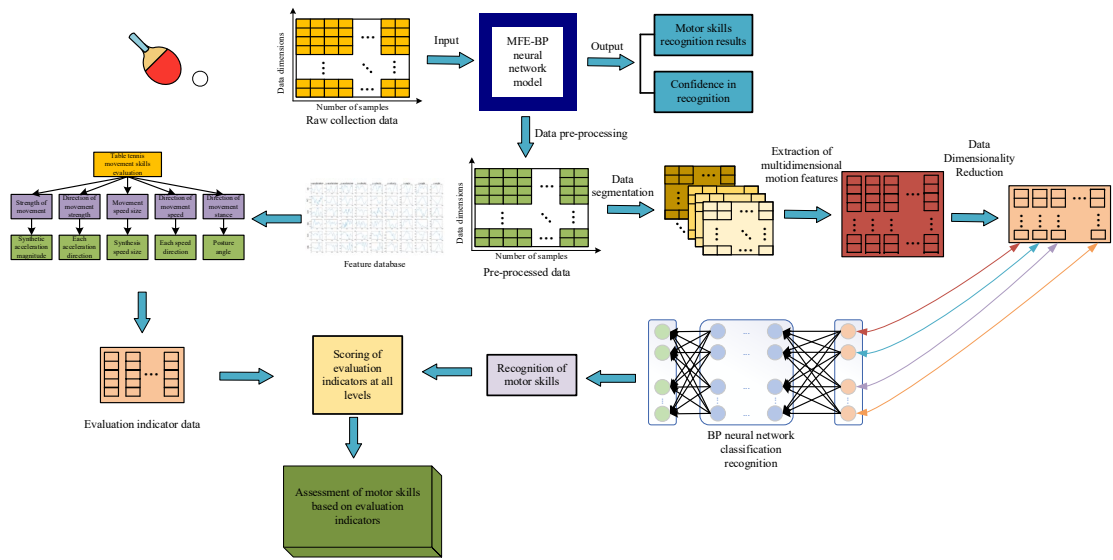

Figure 8 Schematic Diagram of Table Tennis Motor Skill Recognition and Evaluation Algorithm

## 3.1 Motion feature extraction based on feature engineering

The collected players' movement data is transmitted in the way of data stream, which cannot be processed by feature engineering directly. Therefore, the sliding window function in 2.1 is used to process the collected movement data, and then the movement features are extracted and collected based on feature engineering.

The collected movement information is shown in Table 1, including triaxial acceleration $<acc\_x_i, acc\_y_i, acc\_z_i>$, triaxial angular velocity data $<\omega\_x_i, \omega\_y_i, \omega\_z_i>$ and triaxial posture data $<\theta\_x_i, \theta\_y_i, \theta\_z_i>$. After preprocessing and segmentation of the collected data, movement features are extracted respectively according to the acceleration, angular velocity and posture in the three directions XYZ. Common features include mean value, variance and other features that reflect the original collected data samples. However, the dimensional characteristic values changes due to different some external physical quantities, and cannot reflect the original detailed features of the waveform, which cuases certain difficulties to engineering applications. Therefore, we need to adopt more dimensionless motion features to describe the motion information. It can also provide more comprehensive motion features for motion pattern recognition. Based on this, the motion features extracted in this paper are as follows.

The extracted motion features including dimension include: the mean value of the motion information in XYZ direction and the resultant direction, the variance of the motion information in XYZ direction and the resultant direction, the maximum and minimum values of the motion information in XYZ direction and the resultant direction, the peaking and valley values of the motion information in XYZ direction and the resultant direction, the mean square value of the motion information in XYZ direction and the resultant direction, the root mean square of the XYZ direction and the resultant direction.

Motion features without dimensional effects include: crest factor of motion information in XYZ direction and resultant direction, pulse factor of motion information in XYZ direction and resultant direction, margin factor of motion information in XYZ direction and resultant direction, kurtosis factor of motion information in XYZ direction and resultant direction, waveform factor of motion information in XYZ direction and resultant direction, and skewness of motion information in XYZ direction and resultant direction.

Taking a group of data after using the window as an example, the extracted features and corresponding calculation methods are shown in Table 2.

Table 2 The Extracted Motion Features of Players

| Feature Name | Feature Explanation | Calculation Equation | |
|---|---|---|---|
| Mean Value | The average level of the average factor in signal | $\bar{x} = \frac{1}{n}\sum_{i=1}^{n} x_i$ | (14) |
| Variance | The intensity of the factor fluctuation in signal | $\sigma_x^2 = \frac{1}{n}\sum_{i=1}^{n}(x_i - \bar{x})^2$ | (15) |
| Maximum Value | The maximum value of the factor in signal | $\max\_x = \max(x_i)$ | (16) |
| Minimum Value | The minimum value of the factor in signal | $\min\_x = \min(x_i)$ | (17) |
| Peak-valley Value | The range of fluctuation of a factor in signal | $x_{pv} = \max(x_i) - \min(x_i)$ | (18) |

| Mean Square | Signal energy | $\psi_x^2 = \frac{1}{n}\sum_{i=1}^{n} x_i^2$ | (19) |
|---|---|---|---|
| Root mean square | The effective value of signal | $x_{rms} = \sqrt{\frac{1}{n}\sum_{i=1}^{n} x_i^2}$ | (20) |
| Correlation Coefficient | The degree of correlation between different directions | $\rho(X,Y) = \frac{\text{cov}(X,Y)}{\sigma_x \cdot \sigma_y}$ | (21) |
| Crest Factor | How extreme the peak is in the waveform | $\frac{\max(x_i) - \min(x_i)}{\sqrt{\frac{1}{n}\sum_{i=1}^{n} x_i^2}}$ | (22) |
| Pulse Factor | How extreme the peak is in the waveform | $\frac{\max(x_i) - \min(x_i)}{\frac{1}{n}\sum_{i=1}^{n} |x_i|}$ | (23) |
| Margin factor | How extreme the peak is in the waveform | $\frac{\max(x_i) - \min(x_i)}{\left(\frac{1}{n}\sum_{i=1}^{n} \sqrt{|x_i|}\right)^2}$ | (24) |
| Kurtosis Factor | The flatness of the waveform | $x_{kr} = \frac{\frac{1}{n}\sum_{i=1}^{n} x_i^4}{\sqrt{\frac{1}{n}\sum_{i=1}^{n} x_i^2}}$ | (25) |
| Waveform Factor | The flatness of the waveform | $\frac{\sqrt{\frac{1}{n}\sum_{i=1}^{n} x_i^2}}{\frac{1}{n}\sum_{i=1}^{n} |x_i|}$ | (26) |
| Skewness | How far a wave crest is off center | $x_{sk} = \frac{\frac{1}{n}\sum_{i=1}^{n}(x_i - \bar{x})^3}{\left(\frac{1}{n}\sum_{i=1}^{n}(x_i - \bar{x})^2\right)^{\frac{3}{2}}}$ | (27) |
| Kurtosis | The steepness of the waveform | $x_{ku} = \frac{\frac{1}{n}\sum_{i=1}^{n}(x_i - \bar{x})^4}{\left(\frac{1}{n}\sum_{i=1}^{n}(x_i - \bar{x})^2\right)^2} - 3$ | (28) |

After data preprocessing, the data is segmented by sliding window function, and 180 features are extracted for data classification.

## 3.2 Feature dimension reduction

In the field of machine learning, overlearning directly has a negative influence on the classification accuracy [41-43] and causes much more burden to the machine learning. Therefore, data dimension reduction is necessary. Principal Component Analysis (PCA) is a typical unsupervised dimension reduction algorithm, which can reduce multiple indicators into several principal components. They are linearly independent and formed by linear combination of original variables, and can reflect most useful information in original data [44,45]. There are 18 dimensions of motion feature data extracted in this study so, it is necessary to adopt the principal component analysis model to reduce the dimension of the extracted motion features so as to construct effective motion features.

A matrix is formed based on the indexes and classification labels in the motion feature data, and the mean value of each element of the matrix is subtracted to get the decentralized matrix, which is shown in the formula.

$$X_{ij} = \begin{bmatrix} x_{11} - \frac{1}{m}\sum_{i=1}^{m} x_{i1} & x_{12} - \frac{1}{m}\sum_{i=1}^{m} x_{i2} & \cdots & x_{1n} - \frac{1}{m}\sum_{i=1}^{m} x_{in} \\ x_{21} - \frac{1}{m}\sum_{i=1}^{m} x_{i1} & x_{22} - \frac{1}{m}\sum_{i=1}^{m} x_{i2} & \cdots & x_{2n} - \frac{1}{m}\sum_{i=1}^{m} x_{in} \\ \vdots & \vdots & \ddots & \vdots \\ x_{m1} - \frac{1}{m}\sum_{i=1}^{m} x_{i1} & x_{m2} - \frac{1}{m}\sum_{i=1}^{m} x_{i2} & \cdots & x_{mn} - \frac{1}{m}\sum_{i=1}^{m} x_{in} \end{bmatrix} \quad (29)$$

The covariance matrix was calculated according to the decentralized matrix of motion feature data, as shown in the equation

$$C = \frac{1}{m-1} X^T X \quad (30)$$

The feature decomposition of the covariance matrix $C$ is carried out to find the eigenvalues of the covariance matrix $\lambda_k$ and the corresponding eigenvectors $v_k$, as shown in the formula.

$$Cv_k = \lambda_k v_k \quad (31)$$

The contribution rate $c_i$ of the $i$-th component to the total component is defined as shown in the equation.

$$c_i = \frac{\lambda_i}{\sum_{k=1}^{p} \lambda_k} (i = 1, 2, \ldots, p) \quad (32)$$

In the above equation, $\lambda_i (i = 1, 2, \ldots, p)$ is the eigenvalue of the covariance matrix $R$.

The definition of cumulative contribution rate is shown in the next equation.

$$C_i = \frac{\sum_{k=1}^{i} \lambda_k}{\sum_{k=1}^{p} \lambda_k} (i = 1, 2, \ldots, p) \quad (33)$$

The cumulative contribution rate is successively accumulated. When the cumulative contribution rate exceeds 95%, it shall be regarded as the principal component of motion features.

In order to improve the accuracy of machine learning, principal component analysis is used to reduce the dimension of the extracted motion features. The information retention rate of the principal component is set as 95%, and there is the new feature with 23 dimensions. The new feature is constructed as a linear combination of the original feature, as shown in the equation.

$$F_n = \sum_{i=1}^{180} k_i \cdot f_i \qquad (34)$$

As shown in the equation, $f_i$ is the $i$-th original feature, $k_i$ is the weight coefficient of the i-th original feature, $F_n$ is the NTH feature constructed.

## 3.3 Motor skill recognition model

Classification is one of the core problems in data mining, machine learning and pattern recognition [46]. Common classification models include discriminant analysis, cluster analysis, support vector machine and neural network. Support vector machine model is to find a hyperplane farthest away from different categories of samples to realize the classification and recognition of data. The neural network model classifies samples by exploring the functional relationship between input and output. This chapter uses the processed data to establish support vector machine model and neural network model respectively for table tennis skill identification and comparison.

3.3.1 Support vector machine model of the motor skills recognition

In the support vector machine model, a classification method is found to separate the data sets, and the corresponding optimization equation is simplified as shown in the equation.

$$\begin{cases} \min \|\omega\|^2 \\ s.t. y_i \left( \omega^T x + b \right) \geq 1 \end{cases} \qquad (35)$$

In the above equation, $\omega$ is interval, $y_i$ is sample label.

In order to improve the ability of support vector machine to classify nonlinear data sets, the kernel function of support vector machine model is set as Gaussian kernel function. The support vector machine model is used to classify players' motion status. However, the classification of player's motion skills is not a binary classification, so the traditional support vector machine model needs to be improved. Considering that there are only six states of action skills of table tennis players, we choose to improve the support vector machine model based on directed acyclic graph to realize the multi-classification problem of action skills recognition of table tennis players. In this paper, six motor skills were classified and identified based on the Directed Acyclic Graph Support Vector Machine (DAGSVM) model. The classification of directed acyclic graph for the recognition of six motor skills of table tennis players is shown in Figure 9.

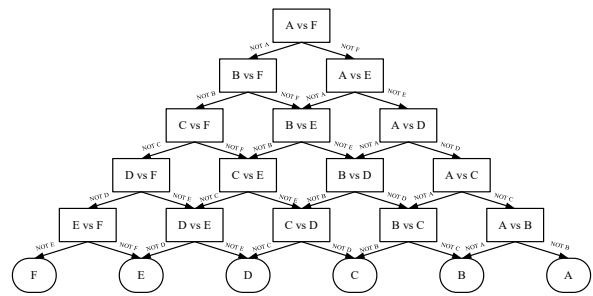

Figure 9 Topology of Directed Acyclic Graphs

As shown in Figure 9, a directed acyclic graph structure based on DAGSVM model can be completed by classifying and identifying 6 motor skills of table tennis players for 5 times.

3.3.2 Motor skill recognition model based on deep learning

Deep learning is a data-driven model, and its internal rules can be fitted by learning sample data [47]. Deep learning is widely used in computer vision, natural language processing and reinforcement learning [48-50]. Based on the new features of dimensionality reduction of data in 3.2, a neural network model is built to recognize and classify motor skills. In this neural network, there are 23 inputs and 6 outputs, and the activation function is softmax function. The input is the 23 new features after dimensionality reduction in 3.2, and the output is to predict the probability of different motor skills corresponding to this group of samples. The output is in the form of one-hot vector. Finally, the classification of maximum probability is selected as the result of motor skill recognition. Parameters for setting the neural network are shown in Table 3, and the corresponding neural network topology is shown in Figure 10.

Table 3 Neural Network Parameter Setting Table

| Parameter Name | Set Value |
| --- | --- |
| The Sample Size of Training Data Set | 60000 |
| The Sample Size of the Test Data Set | 12000 |
| Learning Rate | 0.01 |
| The Size of Input Nodes | 23 |
| The Size of Hidden Layers | 2 |
| The Size of Hidden Nodes | 120 |
| The Size of Output Nodes | 6 |
| Activation Function | Softmax |
| Training Approach | Stochastic Gradient Descent |

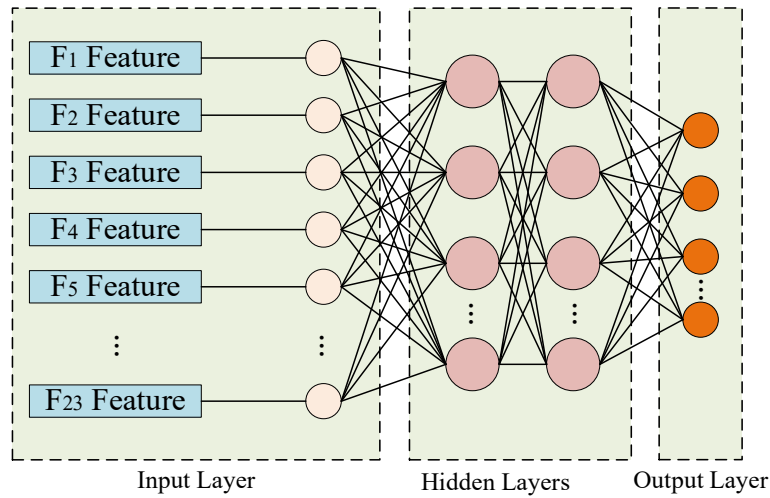

Figure 10 Neural Network Topology

## 4. Motor skill evaluation model

After the recognition of the motor skills of table tennis players, it is necessary to evaluate them. Firstly, choose to establish different evaluation levels; Secondly, determine the relative weights of different evaluation levels. Then, establish different levels of evaluation indicators; Finally, based on the evaluation indexes of different evaluation levels, establish an evaluation model of table tennis players' motor skills.

### 4.1 Hierarchical evaluation system of motor skills

The evaluation system of table tennis motor skills can be comprehensively considered from the following aspects. 1) The fitness of the strength of the motor skill. The strength of a table tennis player in the process of completing a motor skill can be used as an evaluation index to evaluate the player's power, which can be constructed by means of the mean value of the synthetic acceleration. 2) The fitness of direction of playing force. The force direction of table tennis players in completing a motor skill is also an important factor to be considered. This evaluation index can be constructed by three acceleration direction vectors. 3) The fitness of velocity of motor skills. The velocity of a table tennis player in the process of completing a motor skill can measure the velocity of the swing of the bat in playing, and this evaluation index can be constructed by synthesizing the mean value of velocity. 4) The fitness of velocity direction of motor skills. Table tennis players need to consider not only the velocity but also the direction of velocity when completing a motor skill. This index is constructed by three velocity direction vectors. 5) The fitness of posture angle of motor skills. Table tennis players also need to maintain the posture angles corresponding to motor skills in the process of doing a motor skill. This index is constructed by three posture direction vectors. The above five levels can be achieved through the table tennis motor skill evaluation system as shown in Figure 11.

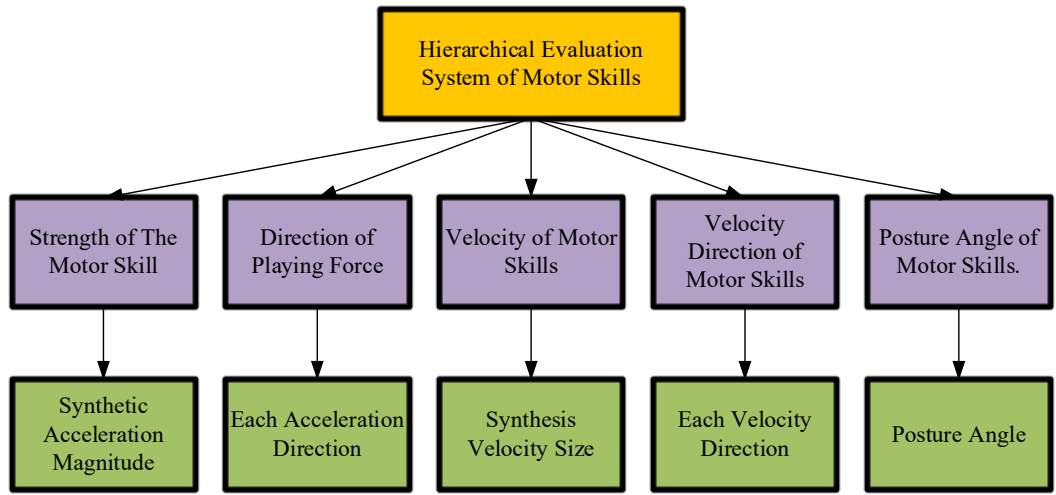

Figure 11 Table Tennis Motor skill Evaluation System

As shown in Figure 11, the evaluation of motor skills is divided into five evaluation levels: strength, force direction, velocity, velocity direction and posture angle. Corresponding evaluation indexes are established for each evaluation level to build an evaluation system. The contribution of the five evaluation levels to the evaluation scores can be written as follows.

$$Q = \sum_{i=1}^{5} K_i \cdot Q_i \tag{36}$$

The weights of different evaluation levels are determined based on analytic hierarchy process, and the pairwise comparison matrix is constructed, as shown in the following.

$$A = \begin{bmatrix} 1 & 1/3 & 1 & 1/5 & 1/7 \\ 3 & 1 & 3 & 3 & 1/5 \\ 1 & 1/3 & 1 & 1/3 & 1/7 \\ 5 & 1/3 & 3 & 1 & 1/3 \\ 7 & 5 & 7 & 3 & 1 \end{bmatrix} \tag{37}$$

Based on the analytic hierarchy process, the shown pairwise comparison matrix is used to calculate the relative weight of five evaluation levels, that is, [0.0556, 0.2055, 0.0592, 0.1715, 0.5081], the scores of five evaluation levels of contribution can be expressed as as shown in the following.

$$Q = 0.056Q_1 + 0.206Q_2 + 0.059Q_3 + 0.172Q_4 + 0.508Q_5 \tag{38}$$

Among the five evaluation levels, the level of strength includes three evaluation indexes of acceleration in three directions. The level of force direction includes three evaluation indexes of acceleration in three directions. The level of velocity includes three evaluation indexes of velocity in three directions. The level of velocity direction includes three evaluation indexes of velocity in three directions. The level of attitude Angle includes three evaluation indexes of Euler Angle in three directions. A level includes three evaluation indicators. The importance of each evaluation indicator in a level is regarded as the same. Based on this, a three-level evaluation system with 15 evaluation indicators is constructed.

## 4.2 Score of each evaluation index

In the same evaluation level, the features of the level are jointly determined by each evaluation index, so we do not consider the difference of the contribution of each evaluation index to the evaluation of the same level. The following is the establishment of evaluation function of each evaluation index.

(1) Level evaluation of strength

In table tennis motor skills, the contribution of acceleration to motor skills is positive, so acceleration should be a very big index. The acceleration data collected are normalized and evaluated by Logistic function, as shown in the equation.

$$Q_1 = 1 - \frac{1}{1 + e^{-\frac{(a - Ea)}{a_{up} - a_{down}}}} \tag{39}$$

In the above, $Ea$ is the average level of acceleration, $a_{up}$ is the maximum value of $a$, $a_{down}$ is the minimum value of $a$. The Logistic function shown in the equation can map the acceleration information into the interval [0,1] to achieve data normalization and thus serve as the evaluation score of the evaluation index.

(2) Level evaluation of force direction

In table tennis motor skills, the contribution of acceleration direction to motor skills should be within the appropriate interval, so the acceleration direction should be an interval index. The interval function is established and normalized for the acceleration data collected, as shown in the equation.

$$Q_2 = \begin{cases} 1 - e^{-\left(\frac{a\_\theta_{down} - a\_\theta}{k_1}\right)} & a\_\theta < a\_\theta_{down} \\ 1 & a\_\theta_{down} < a\_\theta < a\_\theta_{up} \\ 1 - e^{-\left(\frac{a\_\theta - a\_\theta_{up}}{k_2}\right)} & a\_\theta > a\_\theta_{up} \end{cases} \tag{40}$$

In the above, $k_1$ is the loss coefficient lower than the lower boundary. $k_2$ is the loss coefficient higher than the upper boundary. $a\_\theta_{down}$ is the lower bound of the angular direction in the corresponding motor skill. $a\_\theta_{up}$ is the upper bound of the angular direction in the corresponding motor skill. The function shown in the equation can be used to normalize the interval index of the direction of force, and make evaluation scores.

(3) velocity level evaluation

In table tennis motor skills, the contribution of velocity to motor skills is positive, so velocity should be a very big index. The velocity data collected is normalized and evaluated by Logistic function, as shown in the equation.

$$Q_3 = 1 - \frac{1}{1 + e^{\frac{(v - Ev)}{v_{up} - v_{down}}}} \tag{41}$$

In the above, $Ev$ is the average level of velocity, $v_{up}$ is the maximum value of $v$, and $v_{down}$ is the minimum value of $v$. The Logistic function can map the velocity information to the interval [0,1] to achieve the normalization of the velocity data and thus serve as the evaluation score of the evaluation index.

(4) Level evaluation of velocity direction

In table tennis motor skills, the contribution of velocity direction to motor skills should be within the appropriate interval, so velocity direction should be an interval index. The interval function is established and normalized for the collected velocity data, as shown in the equation.

$$Q_4 = \begin{cases} 1-e^{-\left(\frac{v\_\theta_{down}-v\_\theta}{k_1}\right)} & v\_\theta < v\_\theta_{down} \\ 1 & v\_\theta_{down} < v\_\theta < v\_\theta_{up} \\ 1-e^{-\left(\frac{v\_\theta-v\_\theta_{up}}{k_2}\right)} & v\_\theta > v\_\theta_{up} \end{cases} \quad (42)$$

In the above, $k_1$ is the loss coefficient when the velocity is lower than the lower boundary. $k_2$ is the loss coefficient when the velocity is higher than the upper boundary. $v\_\theta_{down}$ is the lower bound of the angular direction in the corresponding motor skill. $v\_\theta_{up}$ is the upper bound of the angular direction in the corresponding motor skill. The function shown in the equation can be used to normalize the interval index of velocity direction and make evaluation score.

(5) Posture angle level evaluation

In table tennis action skills, the posture angle direction is the most important to the standard degree of action skills, and the attitude Angle should be within the appropriate interval, so the attitude Angle direction should be an interval index. The interval function is established and normalized for the collected attitude data, as shown in the equation.

$$Q_5 = \begin{cases} 1-e^{-\left(\frac{\alpha_{down}-\alpha}{k_1}\right)} & \alpha < \alpha_{down} \\ 1 & \alpha_{down} < \alpha < \alpha_{up} \\ 1-e^{-\left(\frac{\alpha-\alpha_{up}}{k_2}\right)} & \alpha > \alpha_{up} \end{cases} \quad (43)$$

In the above, $k_1$ is the loss coefficient when the velocity is lower than the lower boundary. $k_1$ is the loss coefficient when the velocity is higher than the upper boundary. $v\_\theta_{down}$ is the lower bound of the angular direction in the corresponding motor skill. $v\_\theta_{up}$ is the upper bound of the angular direction in the corresponding motor skill. The function shown in the equation can be used to normalize the interval index of velocity direction and make evaluation score.

# 5. Results and analysis

The artificial intelligence player motor skill recognition method and motor skill hierarchical evaluation system constructed by the features established in this paper are calculated, and the model results are solved and analyzed.

## 5.1 Results of motor skill identification of table tennis players

In order to solve the table tennis player motor skill recognition model based on feature construction proposed in this paper, the motor skill recognition model based on feature construction support vector machine model and the motor skill recognition model based on feature construction neural network model are solved respectively and compared with the traditional convolutional neural network.

The training performance of the feature-based neural network model and the traditional convolutional neural network model is shown in Figure 12.

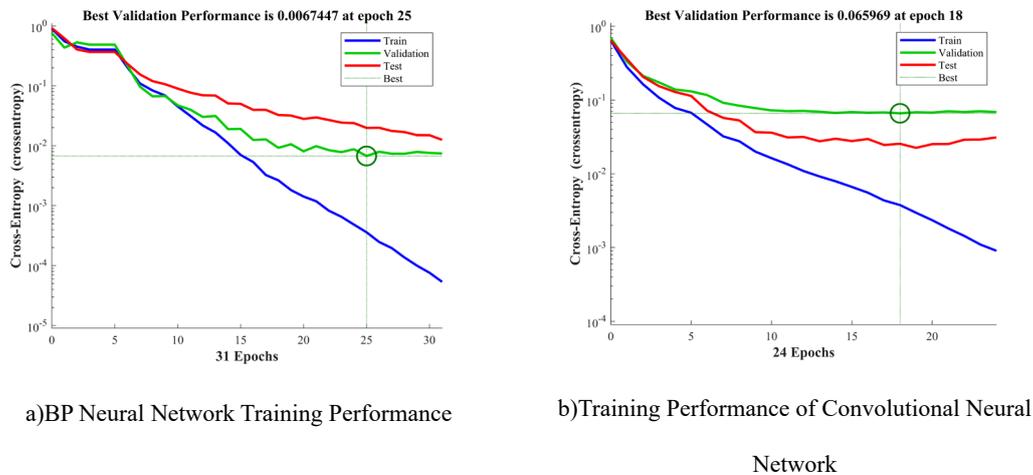

a) BP Neural Network Training Performance

b) Training Performance of Convolutional Neural Network

Figure 12 Training Performance of Two Neural Network Models

As shown in Figure 12 , it can be seen that BP neural network based on feature construction is superior to traditional convolutional neural network in training data set and test data set, and the error of BP neural network based on feature construction is smaller.

The support vector machine model based on feature construction is respectively compared. The confusion matrix of the neural network model based on feature construction and the traditional convolutional neural network model is shown in Figure 13.

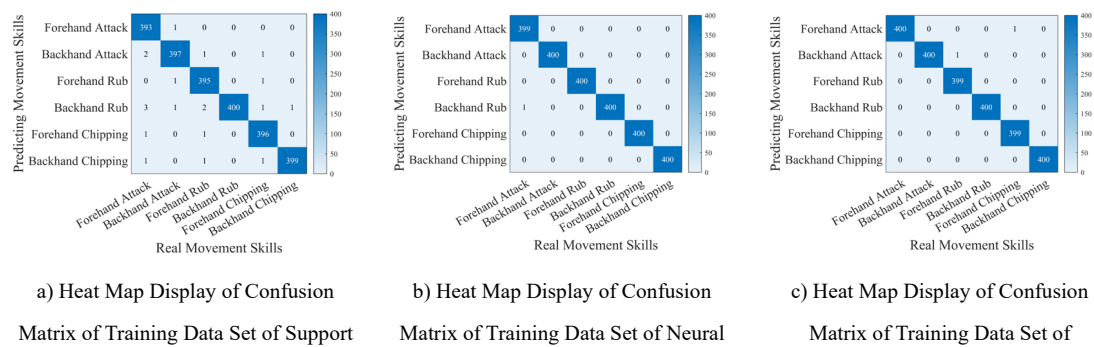

a) Heat Map Display of Confusion Matrix of Training Data Set of Support

b) Heat Map Display of Confusion Matrix of Training Data Set of Neural

c) Heat Map Display of Confusion Matrix of Training Data Set of

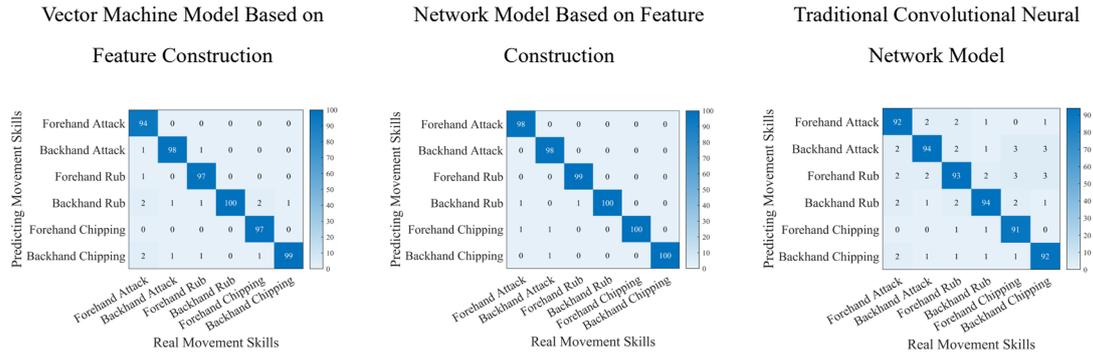

Figure 13 Performance of the Three Prediction Models in the Training Data Set and the Test Data Set

In the classification model, some indicators are commonly used to measure the classification effect of the classification model, such as precision, recall, F1 measure, etc. Precision describes the classification accuracy of the classifier, recall reflects the completeness of the classifier's search for a certain real category, and the F1 measure reflects the robustness of the model, considering precision and recall comprehensively, .

The precision of any classification in the classifier is defined as the proportion of the real category of the classification in all predictions of the classification. The precision is shown in the formula.

$$P = \frac{TP}{TP+FP} \times 100\% \qquad (44)$$

In the above equation, $TP$ is the number of real categories in the category. $FP$ is the number of predicted categories in the category which real categories are not in.

The recall of any classification in the classifier is defined as the proportion of the real category of the classification in all the real categories, and the recall is shown in the equation.

$$P = \frac{TP}{TP+FN} \times 100\% \qquad (45)$$

In the above equation, $TP$ is the number of real categories in the category. $FN$ is the number of real categories in this category which predicted categories are not in.

The F1 measure defining any classification in the classifier is the proportion of the real category of the classification to the linear combination of the predicted category and the real category. The F1 measure is shown in the following.

$$P = \frac{2TP}{2TP + 2\alpha FN + 2(1-\alpha)FP} \times 100\% \qquad (46)$$

In the above, $TP$ is the number of real categories in the category. $FP$ is the number of predicted categories in the category which real categories are not in. $FN$ is the number of real categories in the category but predicted categories are not in. $\alpha$ is the weight coefficient of precision and recall, the value interval is [0,1], which is used to determine the preference of precision and recall in the measurement of the final classifier.

In the recognition of motor skills of table tennis players, the accuracy of real classification recognition is more sensitive, that is, the weight of precision should be greater than the recall. Therefore, the weight coefficient of precision and recall is set as $\alpha = 0.7$, and F1 measure is used to measure the classification ability of the model.

By comparing the support vector machine model based on feature construction, the F1 measure scores of the neural network model based on feature construction and the traditional convolutional neural network model on the training data set and the test data set are shown in Figure 14.

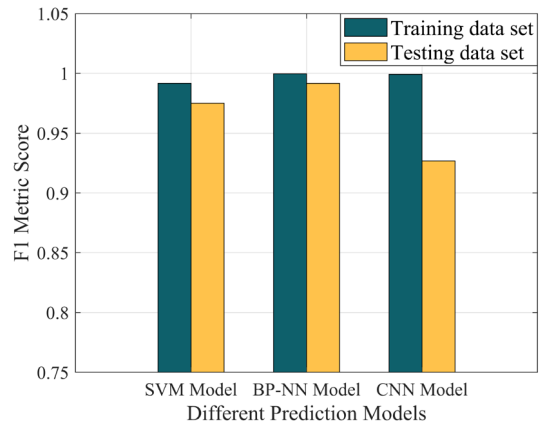

Figure 14 F1 Scores of the Three Prediction Models on the Training Data Set and the Test Data Set

As shown in Figure 14, in the training data set, classification effects are respectively: feature-constructed neural network model, traditional convolutional neural network and feature-constructed support vector machine model. In the test data set, the classification effects are as follows: feature constructed neural network model, feature constructed support vector machine model and traditional convolutional neural network. When extending the model learned from the training data set to the test data set, the recognition effect of the support vector machine (SVM) model and the BP neural network (BP-NN) model declines a bit, but the decline trend of the recognition effect of the traditional convolutional neural networks (CNN) model is more obvious, which indicates that the traditional CNN model has the phenomenon of local overfitting, and the generalization ability of the model is not as good as the BP neural network model constructed based on the feature.

## 5.2 Evaluation of system results

Based on the evaluation system in 4.2, the feature database of standard players' movements is established, and the evaluation model is established with the evaluation criteria of movements in the feature database. The movement data of 50 professional players and 50 amateur players are respectively uploaded for scoring. The scoring results are shown in Figure 15.

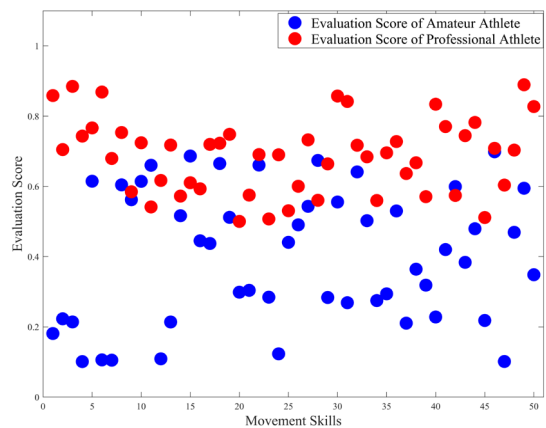

Figure 15 Scoring Diagram of Evaluation Model

As shown in Figure 15, on the whole, professional players score better on motor skills than amateur players. The scores of 50 motor skills of professional and amateur players are counted respectively, as shown in Figure 16.

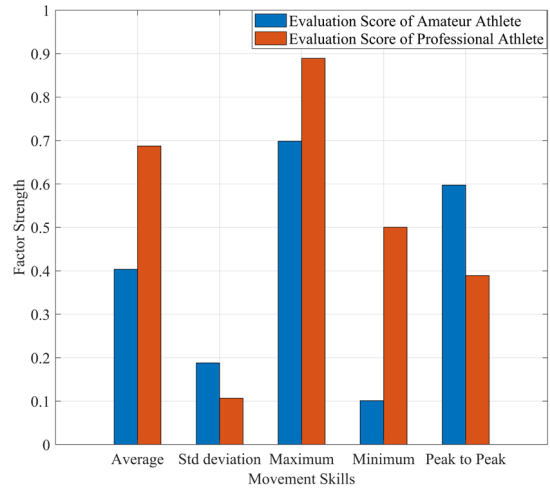

Figure 16 Table tennis players score statistics

As shown in Figure 16, the average score of professional players is higher than that of amateur players, the fluctuation of motor skill score is lower than that of amateur players, the maximum value of motor skill score is higher than that of amateur players, the minimum value of motor skill score is higher than that of amateur players, and the range of motor skill score is smaller than that of amateur players. It can be seen from Figure 16 that the motor skill evaluation system can make scientific evaluation on the motor skills of table tennis players and play a reasonable role in evaluating the motor skills of table tennis players.

# 6 Conclusion

Based on the use of various motor skills in table tennis and the need of monitoring the standardization of motor skills, this study established a set of motor skills recognition and evaluation system for table tennis players. The system includes movement information collection, movement information processing, motor skill recognition, motor skill evaluation and other functions of table tennis players. The main contributions of this study are as follows:

1) Developed a set of table tennis player data measurement system, which can be used to measure player' movement information.

2) The characteristic database of six standard motor skills of table tennis players is established, which is used to establish the standard motor skills.

3) The motor skill recognition model based on the movement information of table tennis players is established, which can realize the recognition of the motor skill of table tennis players.

4) The hierarchical evaluation system of table tennis players' motor skills has been established, which can evaluate the corresponding motor skills of table tennis players.